\documentclass[10pt,conference]{IEEEtran}  %draftcls,onecolumn
\usepackage{amsmath,amsthm, epsfig, verbatim, color, amsfonts} % spconf
\usepackage{graphicx}
\usepackage[pdftex]{hyperref} % turns all linked elements into hypertext
\title{An Achievable Rate for the MIMO Individual Channel}
\author{Yuval Lomnitz, Meir Feder \\
Tel Aviv University, Dept. of EE-Systems  \\
Email: \{yuvall,meir\}@eng.tau.ac.il}

\theoremstyle{plain}
\newtheorem{theorem}{Theorem}
\newtheorem{lemma}{Lemma}%[theorem] % enable the remark in order to number the lemmas as subnumbering of theorems

\theoremstyle{definition}
\def\vr{\mathbf}
\def\mt{\mathbf}

\def\cov{\text{cov}}

\def\Pr{\mathrm{Pr}}
\def\Normal{\mathcal{N}}

\def\Remp{R_{\mathrm{emp}}}

\def\half{\frac{1}{2}}

\def\endofproof{\hspace{\stretch{1}}$\Box$}
\def\doubleint{\int \hspace{-1.5ex} \int}

\newcommand{\argmax}[1] {\raisebox{-1.2ex}{$\stackrel{\textstyle \mathrm{argmax}}{\scriptscriptstyle #1}$}}

\newcommand{\placeunder}[2] {\raisebox{-1.2ex}{$\stackrel{\textstyle #1}{\scriptscriptstyle #2}$}}

{\hspace{\stretch{1}} \textcolor{red}{$\clubsuit^{-1}$ }}
\allowdisplaybreaks[1]
\begin{document}
\maketitle

\begin{abstract}
We consider the problem of communicating over a multiple-input multiple-output (MIMO) real valued channel for which no mathematical model is specified, and achievable rates are given as a function of the channel input and output sequences known a-posteriori. This paper extends previous results regarding individual channels by presenting a rate function for the MIMO individual channel, and showing its achievability in a fixed transmission rate communication scenario.
\end{abstract}

%\begin{IEEEkeywords}
%\end{IEEEkeywords}

\IEEEpeerreviewmaketitle

\section{Introduction}
We consider a channel, termed an \textit{individual channel}, where no specific probabilistic or mathematical relation between the input and the output is assumed. This channel is an extreme case of an unknown channel. Achievable rates are characterized using only the input and output sequences, which capture the actual (a-posteriori) channel behavior. This point of view is similar to the approach used in universal source coding of individual sequences where the goal is to asymptotically attain for each sequence the same coding rate achieved by the best encoder from a model class, tuned to the sequence. This framework is an evolution of those considered in Shayevitz and Feder \cite{Ofer_EMP} and Eswaran et. al. \cite{Eswaran} and is presented in more detail in our papers \cite{Individual_channels_full} and \cite{Individual_channels_isit}, together with the relevant background. We will give a brief introduction below. %\cite{Ofer_BSC} \cite{Eswaran_conf}

The setting we consider includes a single encoder receiving a message to transmit and emitting symbols $x_i \in \mathcal{X}, i=1,2,..,n$ and a decoder receiving a sequence of symbols $y_i \in \mathcal{Y}, i=1,2,..,n$ and attempting to reconstruct the message. In the present paper the input and output symbols are real-valued vectors, i.e. $\mathcal{X} = \mathbb{R}^t$ and $\mathcal{Y} = \mathbb{R}^r$. The relation between $\vr x = [x_1,...,x_n]^T$ and $\vr y = [y_1,...,y_n]^T$ is unknown to the encoder and decoder. We consider two communication scenarios: with feedback (possibly imperfect) and without feedback. For the case in which there is no feedback the communication system transmits in a constant rate, and outage is unavoidable, i.e. one cannot guarantee a small probability of error in all circumstances. In the case feedback exists, the communication rate may be dynamically adapted and outage may be prevented. In both cases we assume common randomness exists in the encoder and the decoder. The results in the current paper extend the previous results, yet only for the first case, of transmission in a constant rate.

The performance is measured by a rate function $\Remp : \mathcal{X}^n \times \mathcal{Y}^n \to \mathbb{R}$ representing an empirical measure of the achievable rate between the channel input and channel output, over $n$ channel uses. In examples here and in \cite{Individual_channels_full}\cite{Individual_channels_isit}, $\Remp$ can be viewed as the mutual information achieved in a certain family of statistical models (in the current scope, all zero mean Gaussian channels), when the model parameters match the empirical ones. In communication without feedback we say that a given rate function $\Remp(\vr x, \vr y)$ is achievable with an input distribution $Q(\vr x)$ if for large block size $n \rightarrow \infty$, it is possible to communicate at any rate $R$ and an arbitrarily small error probability is obtained whenever $\Remp(\vr{x}, \vr{y}) > R$.
%\textcolor{blue}{In communication with feedback we say a given rate function is achieved by a communication scheme if for large block size $n$, data at rate close to or exceeding $\Remp(\vr{x}, \vr{y})$ is decoded successfully with arbitrarily large probability for every output sequence and almost every input sequence. In both cases}
The communication system is required to emit blocks with probability distribution $Q(\vr x)$, which is possible due to the use of randomization. By placing this additional constraint we leave aside the question of adapting the input distribution, so that the current framework attempts at achieving the empirical "mutual information" rather than the empirical "capacity". Another reason for the fixed prior is avoiding degenerate systems which may transmit only "bad" sequences with low (or zero) $\Remp$. This constraint is further discussed in \cite{Individual_channels_full}, section VIII.C.

The main result of this paper is that for the multiple-input multiple-output (MIMO) channel $\mathbb{R}^t \to \mathbb{R}^r$ (i.e. with $t$ transmit and $r$ receive antennas) the rate function defined below is asymptotically achievable, in the fixed rate case:
\begin{equation}\label{eq:Remp_MIMO}
\Remp(\mt X, \mt Y) = \frac{1}{2} \log \left( \frac{ \left|  \hat {\mt R}_{XX} \right| \cdot  \left| \hat {\mt R}_{YY} \right|}{ \left| \hat {\mt R}_{(XY)(XY)} \right|} \right)
\end{equation}
where the $n \times t$ matrix $\mt X$ denotes the channel input over $n$ symbols, and the $n \times r$ matrix $\mt Y$ denotes the output. $\hat {\mt R}_{XX} = \frac{1}{n} \mt X^T \mt X$,  $\hat {\mt R}_{YY} = \frac{1}{n} \mt Y^T \mt Y$ and $\hat {\mt R}_{(XY)(XY)} = \frac{1}{n} [\mt X \mt Y]^T [\mt X \mt Y]$ are the input, the output and the joint empirical correlation matrices, respectively. This is a generalization of the result of \cite{Individual_channels_full} where the rate function $\Remp = \half \log \left( \frac{1}{1-\hat\rho(\vr x, \vr y)^2} \right)$ was proved to be achievable for real valued SISO channel $\mathbb{R} \to \mathbb{R}$ ($\hat\rho$ denotes empirical correlation). As in \cite{Individual_channels_full}, the proof is geometrically intuitive. The results easily extend to the \textit{complex} MIMO case, and to rate function using the empirical \textit{covariance} (rather than the correlation), but we focus here on the simpler case.

%As demonstrated in \cite{Individual_channels_full}, achievability results for individual channels, combined with the law of large numbers, can be used to show various achievability results for probabilistic, and semi-probabilistic models. Thus, this result can be used not only to show existing results such as the attainability of the capacity of the MIMO channels, but also new results. For example, we can extend Lapidoth's result \cite{Lapidoth_Nearest} that the Gaussian capacity $\half \log (1+\textit{SNR})$ can be obtained by the same receiver for any noise distribution, to the MIMO case.

The paper is organized as follows: in Section \ref{sec:Rate_function_origin} we explain the motivation for this rate function and its relation to the probabilistic Gaussian channel, in Section \ref{sec:main_result} we present in detail the main result, which is proven in Section \ref{sec:proof_of_lemma}. Section \ref{sec:comments} is devoted to comments and further research items.

We use lowercase boldface letters to denote vectors, and uppercase boldface letters to denote matrices. We use the same notation for random variables and their sample values, and the distinction should be clear from the context. %$\vr x_i^j$ is used to denote the subvector containing elements from $i$ to $j$ of $\vr x$.

\section{Origin of the rate function} \label{sec:Rate_function_origin}
Consider the channel from $\vr x \in \mathbb{R}^t$ to $\vr y \in \mathbb{R}^r$ which are real valued vectors. For the additive white Gaussian noise (AWGN) MIMO channel $\vr y = \mt H \vr x + \vr v$ with $\vr v \sim \Normal(0,\sigma^2 \mt I)$, and $\vr x \sim \Normal(0,\mt I)$ it is well known that the mutual information is
\begin{equation}\label{eq:MI_MIMO_AWGN}
I(\vr x; \vr y) = \half \log \left| \mt I + \frac{1}{\sigma^2} \mt H^T \mt H \right|
\end{equation}
see for example \cite{Telatar_MIMO}\cite{Goldsmith03}. This reflects the maximum achievable rate with the fixed covariance matrix $E \vr x \vr x^T = \mt I$, and is sometimes termed the \textit{open-loop MIMO capacity}, since equal power is a reasonable choice when the transmitter does not know the channel. A more general form of the mutual information is obtained by assuming $\vr x, \vr y$ are any jointly Gaussian random vectors and writing:
\begin{eqnarray}
h(\vr x) &=& \half \log | 2 \pi e \cdot \cov(\vr x) | \\
h(\vr y) &=& \half \log | 2 \pi e \cdot \cov(\vr y) | \\
h(\vr x, \vr y) &=& \half \log \left| 2 \pi e \cdot \cov \left( \left[ \begin{array}{c} \vr x \\ \vr y \end{array} \right]\right) \right|
\end{eqnarray}
Therefore:
\begin{equation}\label{eq:MI_MIMO}
I(\vr x;\vr y) = h(\vr x) + h(\vr y) - h(\vr x, \vr y) = \half \log \left[ \frac{| \cov(\vr x) | \cdot | \cov(\vr y) |}{\left| \cov \left( \left[ \substack{ \vr x \\ \vr y }  \right] \right) \right|} \right]
\end{equation}
where the factors $2 \pi e$ cancel out since the dimension of the covariance matrix in the denominator is the sum of the dimensions in the nominator. The expression (\ref{eq:MI_MIMO}) is more general than (\ref{eq:MI_MIMO_AWGN}) since it does not assume the noise is white, and is suitable for our purpose since it expresses the mutual information through properties of the input and output vectors without using an explicit channel structure. For the case of the AWGN MIMO channel it yields the same value as (\ref{eq:MI_MIMO_AWGN}). For the particular scalar case where $\vr x$, $\vr y$ are scalars with variances $\sigma_X^2$, $\sigma_Y^2$ and correlation factor $\rho$, Equation (\ref{eq:MI_MIMO}) evaluates to $I(\vr x;\vr y) = \half \log \left( \frac{1}{1-\rho^2} \right)$, as previously obtained for the SISO case.

The empirical rate function we defined in (\ref{eq:Remp_MIMO}) is an empirical version of the mutual information expression in (\ref{eq:MI_MIMO}), except that the covariance matrices are replaced by empirical \textit{correlation} (rather than covariance) matrices, i.e. we do not cancel the mean. When $|\hat {\mt R}_{XX}|=0$ or $|\hat {\mt R}_{YY}|=0$ (which leads also to $|\hat {\mt R}_{(XY)(XY)}|=0$) , the rate function will be defined by removing the columns of $\mt X$ or $\mt Y$ (respectively), which are linearly dependant on the others, until these determinants become positive. It is not important which columns are removed to break the linear dependence, due to this function's invariance to linear transformation (Property \ref{property2} below). For the case of $\mt Y=0$ or $\mt X=0$ we define $\Remp=0$.

%\begin{multline}
%I(\vr x;\vr y) = \half \log \left[ \frac{\sigma_X^2 \cdot \sigma_Y^2 }{\left| \begin{array}{cc} \sigma_X^2  & \sigma_X \sigma_Y \rho \\  \sigma_X \sigma_Y \rho & \sigma_Y^2  \end{array}  \right|} \right]
%=\\=
%\half \log \left[ \frac{\sigma_X^2 \cdot \sigma_Y^2}{\sigma_X^2 \sigma_Y^2 (1-\rho^2) } \right]
%= \half \log \left[ \frac{1}{1-\rho^2} \right]
%\end{multline}

The rate function has the following properties which are expected from an empirical metric of the ``mutual information'':
\begin{enumerate}\label{sec:MIMO_u0_Remp_properties}
\item \label{property1} \textbf{Non-negativity:} $\Remp(\mt X, \mt Y) \geq 0$. This is evident from the fact $\Remp(\mt X, \mt Y)$ is the mutual information between two Gaussian vectors with the respective covariances. It will also be shown in passing as part of the derivation in Section \ref{sec:proof_of_lemma}.
\item \label{property2} \textbf{Invariance under linear transformations:} Any invertible linear matrix operation on the input or output (for example, multiplying any of the input or output signals by a factor, adding signals, etc) does not change $\Remp(\mt X, \mt Y)$, i.e. $\Remp(\mt X \mt G_x, \mt Y \mt G_y) = \Remp(\mt X, \mt Y)$.

\textit{Proof:} Suppose we multiply $\mt X$ and $\mt Y$ by arbitrary matrices $G_{x,t \times t}$ and $G_{y,r \times r}$ respectively. Define $\mt X' = \mt X \mt G_x$ then $\left| \hat {\mt R}_{XX}' \right| = \left| \frac{1}{n} \mt X'^T \mt X' \right| = \left| \mt G_x \hat {\mt R}_{XX} \mt G_x \right| = \left| \hat {\mt R}_{XX}  \right| \cdot \left| \mt G_x \right|^2$. And likewise for $\mt Y$. Since $[\mt X', \mt Y'] = [\mt X, \mt Y] \cdot \left[ \begin{array}{cc} \mt G_x & 0 \\ 0 & \mt G_y \end{array} \right]$ then from the same considerations we will have $\left| \hat {\mt R}_{(XY)(XY)}' \right| = \left| \hat {\mt R}_{(XY)(XY)}  \right| \cdot \left| \begin{array}{cc} \mt G_x & 0 \\ 0 & \mt G_y \end{array} \right|^2 =  \left| \hat {\mt R}_{(XY)(XY)}  \right| \cdot \left| \mt G_x \right|^2 \cdot \left| \mt G_y \right|^2 $, therefore the factors cancel out and $\Remp(\mt X', \mt Y') = \Remp(\mt X, \mt Y)$

\item \label{property3} \textbf{Symmetry:} $\Remp(\mt X, \mt Y) = \Remp(\mt Y, \mt X)$
\end{enumerate}

\section{The main result} \label{sec:main_result}
\begin{theorem}[Non-adaptive, continuous MIMO channel]\label{theorem:continuous_nonadaptive_MIMO1}
Given the channel $\mathbb{R}^t \to \mathbb{R}^r$, define the input over $n$ symbols as an $n \times t$ matrix $\mt X$, and the output as an $n \times r$ matrix $\mt Y$. Let $\hat {\mt R}_{XX} = \frac{1}{n} \mt X^T \mt X$,  $\hat {\mt R}_{YY} = \frac{1}{n} \mt Y^T \mt Y$ and $\hat {\mt R}_{(XY)(XY)} = \frac{1}{n} [\mt X \mt Y]^T [\mt X \mt Y]$ be the input, the output and the joint empirical correlation matrices, respectively. Define the rate function
\begin{equation}
\Remp(\mt X, \mt Y) = \frac{1}{2} \log \left( \frac{ \left|  \hat {\mt R}_{XX} \right| \cdot  \left| \hat {\mt R}_{YY} \right|}{ \left| \hat {\mt R}_{(XY)(XY)} \right|} \right)
\end{equation}
Then for every $P_e>0$, a positive definite $t \times t$ matrix $\Lambda_x$ and $n \geq t+r$ there exists random encoder-decoder pair of rate $R$ over block size $n$, such that the distribution of the input sequence is $\mt{X} \sim \Normal^n(0,\Lambda_x)$ and for any $\gamma < 1 - \frac{t+r-1}{n}$ the probability of error for any message given an input sequence $\mt{X}$ and output sequence $\mt{Y}$ is not greater than $P_e$ if:
\begin{equation}\label{eq:cond_on_R_to_fall_below_Pe}
R \leq \gamma \cdot \Remp(\mt X, \mt Y) + \frac{\log(P_e)}{n} - t \lceil r/2 \rceil \frac{\log(n)}{n} - \frac{\log(C_L)}{n}
\end{equation}
where
\begin{equation}\label{eq:value_of_C_L}
C_L = \frac{1}{\Gamma^{t} \left(\frac{r}{2} \right) 2^{t \lceil r/2 \rceil}} \cdot \left( \frac{2}{(1-\gamma)n-t-r+1} + \frac{2}{r} \right)^{t}
\end{equation}
Specifically, for every $\delta > 0$ and $\gamma < 1$ there exists $n$ large enough so that the probability of error is not greater than $P_e$ if:
\begin{equation}\label{eq:asymptotic_cond_on_R_to_fall_below_Pe}
R \leq \gamma \cdot \Remp(\mt X, \mt Y) - \delta
\end{equation}
\end{theorem}

The theorem almost directly follows from the next lemma which we will prove subsequently:
\begin{lemma}\label{lemma:pairwise_MIMO}
For any $n \times r$ matrix $\mt Y$, the probability of $\Remp(\mt X, \mt Y) \geq T$ where $\mt X$ is randomly drawn $\mt{X} \sim \Normal^n(0,\Lambda_x)$ is bounded by:
\begin{equation}
\Pr \{ \Remp(\mt X, \mt Y) \geq T \} \leq C_L \cdot n^{t \lceil r/2 \rceil} \exp(-\gamma \cdot n \cdot T)
\end{equation}
For any $\gamma$ in the range $0 \leq \gamma < 1 - \frac{t+r-1}{n}$, and where $C_L$ is defined in (\ref{eq:value_of_C_L}).
\end{lemma}
Note that the bound does not depend on $\Lambda_x$. To prove Theorem \ref{theorem:continuous_nonadaptive_MIMO1}, the codebook $\{\mt X_m\}_{m=1}^{\exp(nR)}$ is randomly generated by i.i.d. selection of each codeword from the Gaussian matrix distribution $\Normal^n(0,\Lambda_x)$. The common randomness is the codebook itself. The encoder sends the $w$-th codeword, and the decoder uses maximum empirical rate decoder i.e. chooses:
\begin{equation}
\hat{w} = \argmax{m} \{ \Remp(\mt X_m ; \mt Y) \}
\end{equation}
where ties are broken arbitrarily. By using Lemma \ref{lemma:pairwise_MIMO} and the union bound, the probability of error given $\mt X_w, \mt Y$ is bounded by:
\begin{multline}
P_e^{(w)}(\mt X_w, \mt Y)
\leq \\ \leq
\Pr \left\{ \bigcup_{m \neq w} \left( \Remp (\mt X_m ; \mt Y) \geq \Remp (\mt X_w ; \mt Y) \right) \Bigg| \mt X_w \right\} \leq \\
\leq \exp(nR) \cdot C_L \cdot n^{t \lceil r/2 \rceil} \exp(-\gamma \cdot n \cdot \Remp (\mt X_w ; \mt Y))
=\\=
C_L \cdot n^{t \lceil r/2 \rceil} \exp[n (R - \gamma \cdot \Remp (\mt X_w ; \mt Y))]
\end{multline}

Therefore if (\ref{eq:cond_on_R_to_fall_below_Pe}) is satisfied, then $P_e^{(w)}(\mt X_w, \mt Y) \leq P_e$, which proves the first part of the theorem. The second part follows directly from the first part. For any $\gamma < 1$ and $\delta>0$ there is $n$ large enough so that the condition $\gamma < 1 - \frac{t+r-1}{n}$ is satisfied, and then $n$ could be increased till the redundancy in (\ref{eq:cond_on_R_to_fall_below_Pe}), $\frac{\log(P_e)}{n} - t \lceil r/2 \rceil \frac{\log(n)}{n} - \frac{\log(C_L)}{n}$ would be smaller than $\delta$ (note that $C_L$ is decreasing in $n$), therefore $P_e^{(w)}(\mt X_w, \mt Y) \leq P_e$ will be satisfied if (\ref{eq:asymptotic_cond_on_R_to_fall_below_Pe}) is satisfied.
\endofproof

\section{Proof of Lemma 1} \label{sec:proof_of_lemma}
To prove Lemma \ref{lemma:pairwise_MIMO} we use the Chernoff bound:
\begin{multline}\label{eq:Chernoff_error_bound}
\Pr \{ \Remp(\mt X, \mt Y) \geq T \}
=\\=
\Pr \{ \exp(n \gamma \Remp(\mt X, \mt Y)) \geq \exp(n \gamma T) \}
\leq \\ \leq
\frac{E \exp(n \gamma \Remp(\mt X, \mt Y))}{\exp(n \gamma T)} \equiv L \exp(-n \gamma T)
\end{multline}

To prove the lemma we need to calculate
\begin{equation}
L = E \exp(n \gamma \Remp(\mt X, \mt Y)) = E \left( \frac{ \left|  \hat {\mt R}_{XX} \right| \cdot  \left| \hat {\mt R}_{YY} \right|}{ \left| \hat {\mt R}_{(XY)(XY)} \right|} \right)^{\frac{\gamma \cdot n}{2}}
\end{equation}
where the expected value is taken with respect to $\mt X$. The remainder of this section is devoted to upper bounding $L$. We will first assume that $\Lambda_x = \mt I_{t \times t}$, i.e. $\mt X \sim \Normal^n(0,\mt I)$, and then extend to general $\Lambda_x$.

Define $\mt Z = [\mt Y, \mt X]$. We perform a QR decomposition of $\mt X,\mt Y$ and $\mt Z$ in order to obtain more friendly expressions. As a reminder, QR decomposition of a matrix $\mt A_{n \times k} = \mt Q_{n \times k} \mt R_{k \times k}$ (with $\mt Q^T \mt Q = I$ and $\mt R$ upper triangular) is performed by Gram-Schmidt process. We start from the left column of $\mt A$ and work our way to the last one. At each time we take a column of $\mt A$ and split it to the part which can be represented by a linear combination of the columns to the left of it (equivalently, to the columns of $\mt Q$ already generated), and the "innovation", i.e. the part which is orthogonal to the subspace generated by the previous columns. The vector representing the innovation is normalized, and becomes the respective column of $\mt Q$, and its power becomes the diagonal element in $\mt R$. The coefficients representing the part of the vector which is in the subspace of previous columns become the elements of $\mt R$ above the diagonal. Another important property of QR decomposition is that the determinant of $ \mt A^T \mt A$ can be written in terms of the diagonal elements in $\mt R$:  $\left| \mt A^T \mt A \right| = \left| \mt R^T \mt Q^T \mt Q \mt R \right| = \left| \mt R^T \mt R \right| = \left| \mt R \right|^2 = \prod_{i=1}^k R_{ii}^2$.

Now define the diagonal of the upper triangular matrix in the QR decomposition of the matrices $\mt X$, $\mt Y$ and $\mt Z$ respectively to be the vectors $\vr a$, $\vr b$ and $[\vr c, \vr d]$. I.e. if $\mt X = \mt Q_x \mt R_x$, $\mt Y = \mt Q_y \mt R_y$ and $\mt Z = \mt Q_z \mt R_z$ then $\vr a = \mathrm{diag} (\mt R_x)$, $\vr b = \mathrm{diag} (\mt R_y)$, and $[\vr c, \vr d] = \mathrm{diag} (\mt R_z)$. The lengths of the vectors $\vr c, \vr d$ are $r, t$ respectively, so that they overlap with the columns of $\mt Y$ and $\mt X$ in the matrix $\mt Z$. We have:
\begin{equation}
\frac{ \left|  \hat {\mt R}_{XX} \right| \cdot  \left| \hat {\mt R}_{YY} \right|}{ \left| \hat {\mt R}_{(XY)(XY)} \right|}
=
\frac{ \left|  \frac{1}{n} \mt X^T \mt X \right| \cdot  \left| \frac{1}{n} \mt Y^T \mt Y  \right|}{ \left| \frac{1}{n} \mt Z^T \mt Z  \right|}
=
\frac{ \prod_{i=1}^{t} a_i^2  \prod_{i=1}^{r} b_i^2 }{ \prod_{i=1}^{r} c_i^2  \prod_{i=1}^{t} d_i^2 }
\end{equation}
Note that the $\frac{1}{n}$ factors cancel out because the matrix dimensions are $t$ and $r$ in the nominator and $t+r$ in the denominator. Since the Gram-Schmidt process operates sequentially from the first column to the last, and the first $r$ columns of $\mt Z$ and $\mt Y$ are equal, we will have $\vr b = \vr c$. Therefore we can write:
\begin{equation}\label{eq:R_as_ad}
\frac{ \left|  \hat {\mt R}_{XX} \right| \cdot  \left| \hat {\mt R}_{YY} \right|}{ \left| \hat {\mt R}_{(XY)(XY)} \right|}
= \prod_{i=1}^{t} \left( \frac{a_i}{d_i} \right)^2
\end{equation}

Note that $a_i$ and $d_i$ both relate to the same vector, the $i$-th column of $\mt X$.
%They depend on this column, on $\mt Y$ and $d_i$ depends also on the previous columns of $\mt X$.
The ratio $\frac{a_i}{d_i}$ is the ratio between the innovation of the $i$-th column of $\mt X$ with respect to the subspace spanned by previous columns of $\mt X$ alone (nominator) or these columns together with the columns of $\mt Y$ (denominator). Obviously from this reason $|d_i| \leq |a_i|$ (and therefore $\Remp(\mt X, \mt Y) \geq 0$ - Property \ref{property1}).

The key observation in this derivation is as follows: consider a sequential drawing of the columns of $\mt X$ and calculation of the factors $\frac{a_i}{d_i}$. Since the $i$-th column of $\mt X$ is chosen isotropically and independently of the previous columns, the value of previous columns does not affect the distribution of the innovations $d_i, a_i$ (only the number of dimensions in previous columns does). Using this observation which we will prove subsequently, we would be able to break $L$ represented as the expected value of a product (\ref{eq:R_as_ad}) into a product of expected values (equations (\ref{eq:L_prod_induction})-(\ref{eq:L_as_prod_D})), and the proof is completed by a (tedious) calculation of these expected values.

To show the independence of $a_i,d_i$ in previously drawn values, denote by $\mt X_{m}^{i}$ a matrix including the columns $m$ to $i$ of $\mt X$, and by $\vr x_i$ the $i$-th column. Define a a unitary $n \times n$ matrix $\mt Q$ whose first $i-1$ columns span the subspace spanned by the first $i-1$ columns of $\mt X$, its next $r$ columns extend this subspace to cover also the columns subspace of $\mt Y$, and the next $n-(i-1)-r$ columns complete it to an orthonormal basis. This matrix does not depend on $\mt X_{i}^{t}$ and specifically on the column $i$. We assume that the columns of $\mt Y$ are linearly independent (we will relax this assumption later). Also, in probability one, assuming $n \geq t+r$, the columns of $\mt X_1^{i-1}$ are linearly independent of each other and of the columns of $\mt Y$. To see this, it is easy to show that the projection of each column in any direction orthogonal to the subspace already spanned by previous ones (including $\mt Y$), is also Gaussian therefore has probability $0$ to be $0$, as long as there exists such an orthogonal vector, i.e. the number of previously generated vectors is smaller than $n$.

Now define $\vr z = \mt Q^T \vr x_i$. Since $\vr x_i \sim \Normal(0,\mt I_{n \times n})$ also $\vr z \sim \Normal(0,\mt I_{n \times n})$. The first $i-1$ elements of $\vr z$ represent the projections of $\vr x_i$ to the subspace spanned by previous columns of $\mt X$, and the next $r$ elements represent the projections to the subspace spanned by columns of $\mt Y$. So $a_i^2$ collects the energy of all elements except the first $i-1$, and $d_i^2$ collects the energy of all elements except the first $i-1+r$. To see this formally, in the Gram-Schmidt process the coefficients of the projection of $\vr x_i$ on the subspace spanned by $\mt X_1^{i-1}$ are ${\mt Q_1^{i-1}}^T \vr x_i$, and the projection itself is $\mt Q_1^{i-1} {\mt Q_1^{i-1}}^T \vr x_i$, therefore the innovation is $\vr v_i = \vr x_i - \mt Q_1^{i-1} {\mt Q_1^{i-1}}^T \vr x_i$. Since ${\mt Q_{1}^{i-1}}^T \vr v_i = 0$ and $\mt {Q_{i}^n}^T \vr v_i = \mt {Q_{i}^n}^T \vr x_i$ we have
$a_i^2 = \left\| \vr v_i  \right\|^2 = \left\| \mt Q^T \vr v_i \right\|^2 = \left\| \left[ \begin{array}{c}  {\mt Q_{1}^{i-1}}^T \\ \mt {Q_{i}^n}^T \end{array} \right]  \vr v_i \right\|^2 = \left\| \left[ \begin{array}{c} 0 \\ {\mt Q_{i}^n}^T \vr x_i \end{array} \right]  \right\|^2 = \left\|  \vr z_{i}^n  \right\|^2$
%\begin{multline}
%a_i^2 = \left\| \vr x_i - \mt Q_1^{i-1} {\mt Q_1^{i-1}}^T \vr x_i \right\|^2
%=\\=
%\left\| \mt Q^T \left( \vr x_i - \mt Q_1^{i-1} {\mt Q_1^{i-1}}^T \vr x_i \right) \right\|^2
%=\\=
%\left\| \left[ \begin{array}{c}  {\mt Q_{1}^{i-1}}^T \\ \mt {Q_{i}^n}^T \end{array} \right]  \left( \vr x_i - \mt Q_1^{i-1} {\mt Q_1^{i-1}}^T \vr x_i \right) \right\|^2
%=\\=
%\left\| \left[ \begin{array}{c} {\mt Q_{1}^{i-1}}^T \vr x_i - {\mt Q_{1}^{i-1}}^T \vr x_i \\ {\mt Q_{i}^n}^T \vr x_i - 0 \end{array} \right]  \right\|^2
%=\\=
%\left\| \left[ \begin{array}{c} 0 \\ {\mt Q_{i}^n}^T \vr x_i \end{array} \right]  \right\|^2
%=
%\left\| {\mt Q_{i}^n}^T \vr x_i \right\|^2
%=
%\left\|  \vr z_{i}^n  \right\|^2
%\end{multline}
and similarly, $d_i^2 = \left\| \vr x_i - \mt Q_1^{i-1+r} {\mt Q_1^{i-1+r}}^T \vr x_i \right\|^2 = \left\|  \vr z_{i+r}^n  \right\|^2$. Therefore $a_i, d_i$ are independent of $\mt Y$ and the previous columns of $\mt X$, and can be given by norms over parts of a Gaussian i.i.d. vector of length $n$. Defining
\begin{equation}\label{eq:D_i_definition}
D_i \equiv E \left( \left. \left( \frac{a_i}{d_i} \right)^{\gamma n} \right| \mt X_{1}^{i-1} \right) = E \left( \left( \frac{a_i}{d_i} \right)^{\gamma n} \right)
\end{equation}
Where the equality is due to the independence shown above, we have for any $k=1,2,...,t$:
\begin{multline}\label{eq:L_prod_induction}
E \prod_{i=1}^{k} \left( \frac{a_i}{d_i} \right)^{\gamma n}
=
E \left[ E \left( \left. \prod_{i=1}^{k} \left( \frac{a_i}{d_i} \right)^{\gamma n} \right| \mt X_{1}^{k-1} \right) \right]
=\\=
E \left[ \prod_{i=1}^{k-1} \left( \frac{a_i}{d_i} \right)^{\gamma n} \cdot E \left( \left. \left( \frac{a_k}{d_k} \right)^{\gamma n} \right| \mt X_{1}^{k-1} \right) \right]
=\\=
E \left[ \prod_{i=1}^{k-1} \left( \frac{a_i}{d_i} \right)^{\gamma n} \cdot D_k \right]
=
E \left( \prod_{i=1}^{k-1} \left( \frac{a_i}{d_i} \right)^{\gamma n} \right) \cdot D_k
\end{multline}
Therefore by induction:
\begin{equation}\label{eq:L_as_prod_D}
L = E \left[ \frac{ \left|  \hat {\mt R}_{XX} \right| \cdot  \left| \hat {\mt R}_{YY} \right|}{ \left| \hat {\mt R}_{(XY)(XY)} \right|} \right]^{\frac{\gamma \cdot n}{2}} = E \prod_{i=1}^{t} \left( \frac{a_i}{d_i} \right)^{\gamma n} \stackrel{(\ref{eq:L_prod_induction})}{=} \prod_{i=1}^{t} D_i
\end{equation}

Now we bound $D_i$ (using the previously defined Gaussian vector $\vr z$):
\begin{multline}\label{eq:D_i_derivation}
D_i = E \left( \frac{a_i^2}{d_i^2} \right)^{\gamma n / 2}
=
E \left( \frac{\left\| \vr z_{i}^n  \right\|^2}{\left\|  \vr z_{i+r}^n  \right\|^2} \right)^{\gamma n / 2}
%=\\=
%E \left( \frac{ \sum_{i=i}^n z_i^2 }{ \sum_{i=i+r}^n z_i^2 } \right)^{\gamma n / 2}
=\\=
E \left( 1 + \frac{\sum_{j=i}^{i+r-1} z_j^2}{\sum_{j=i+r}^n z_j^2} \right)^{\gamma n / 2}
=\\ \stackrel{(a)}{=}
\raisebox{-2ex}{\placeunder{E}{\substack{ h \sim \chi^2(r) \\ s \sim \chi^2(n-i-r+1)}}} \left( 1 + \frac{h}{s} \right)^{\gamma n / 2}
=\\=
\doubleint_{0}^\infty \left( 1 + \frac{h}{s} \right)^{\frac{\gamma n}{2}} \frac{h^{\frac{r}{2} - 1} e^{-\frac{h}{2}}}{2^{r/2} \Gamma \left(\frac{r}{2} \right)}
\cdot \frac{s^{\frac{n-i-r+1}{2} - 1} e^{-\frac{s}{2}}}{2^{\frac{n-i-r+1}{2}} \Gamma \left(\frac{n-i-r+1}{2} \right)}  \cdot ds \cdot dh
=\\=
\underbrace{\frac{1}{2^{(n-i+1)/2} \Gamma \left(\frac{r}{2} \right)  \Gamma \left(\frac{n-i-r+1}{2} \right)}}_{c_1}
\cdot \\ \cdot
%\int_{s=0}^\infty \int_{h=0}^\infty \left( 1 + \frac{h}{s} \right)^{\gamma n / 2}  h^{\frac{r}{2} - 1} e^{-h/2}  s^{\frac{n-i-r+1}{2} - 1} e^{-s/2} \cdot ds \cdot dh
%=\\=
%c_1
\doubleint_{0}^\infty \left( s + h \right)^{\frac{\gamma n}{2}}  h^{\frac{r}{2} - 1}   s^{\frac{(1-\gamma)n-i-r-1}{2}} e^{-\frac{s+h}{2}} \cdot ds \cdot dh
=\\ \stackrel{(b)}{=}
c_1 \doubleint_{0}^\infty w^{\frac{\gamma n}{2}}  \left( \frac{1}{v+1} w \right)^{\frac{r}{2} - 1}   \left( \frac{v}{v+1} w \right)^{\frac{(1-\gamma)n-i-r-1}{2}}
\cdot \\ \cdot e^{-w/2} \frac{w}{(v+1)^2} \cdot dw \cdot dv
=
c_1 \underbrace{\int_{w=0}^\infty  w^{\frac{n-i-1}{2}} e^{-w/2} dw}_{c_w}
\cdot \\ \cdot
    \underbrace{\int_{v=0}^\infty \left( \frac{1}{v+1} \right)^{\frac{(1-\gamma)n-i+1}{2}} v^{\frac{(1-\gamma)n-i-r-1}{2}} \cdot dv}_{c_v}
\end{multline}
%Note that in the expression above the main difficulty in bounding it relates to the probability of $s = \sum_{i=i+r}^n z_i^2$ to be near $0$.
where in (a) we used independent Chi-Squared distributed $h,s$, and in (b) we changed variables from $s,h$ to $w=s+h, v=s/h$, with inverse transformation $s=\frac{v}{v+1} w, h=\frac{1}{v+1} w$ and Jacobian $J^{-1} = \frac{\partial w,v}{\partial s,h} = \left| \begin{array}{cc} 1 & 1 \\ 1/h & -s/h^2 \end{array} \right| = \frac{s+h}{h^2} = \frac{(v+1)^2}{w}$. The first integral in the expression above evaluates to:
\begin{equation}\label{eq:w_integral}
c_w
%= \int_{w=0}^\infty  w^{\frac{n-i-1}{2}} e^{-w/2} dw = \int_{t=w/2=0}^\infty  (2t)^{\frac{n-i-1}{2}} e^{-t} 2 dt
%= 2^{\frac{n-i+1}{2}} \int_{t=0}^\infty  t^{\frac{n-i+1}{2} - 1} e^{-t} dt
=  2^{\frac{n-i+1}{2}} \Gamma \left( \frac{n-i+1}{2} \right)
\end{equation}
By definition of $\Gamma(\cdot)$. The second integral behaves like $v^{\frac{(1-\gamma)n-i-r-1}{2}}$ near $v=0$ and like $v^{\frac{(1-\gamma)n-i-r-1}{2} - \frac{(1-\gamma)n-i+1}{2}} = v^{\frac{-r-2}{2}}$ at $v \to \infty$. Therefore it will exist (converge) iff the power of $v$ near $0$ is larger than $-1$ and at $\infty$ is smaller than $-1$. The first condition is $\frac{(1-\gamma)n-i-r-1}{2} > -1 \Rightarrow  (1-\gamma)n > i+r-1 \Rightarrow \gamma < 1 - \frac{i+r-1}{n}$. The other condition always holds since $r>0$. Note that since the power of $\frac{1}{v+1}$ is larger by more than 1 than the power of $v$ it is positive (when the first condition holds). Therefore we can bound:

\begin{multline}\label{eq:v_integral_bound}
c_v
%= \int_{v=0}^\infty \left( \frac{1}{v+1} \right)^{\frac{(1-\gamma)n-i+1}{2}} v^{\frac{(1-\gamma)n-i-r-1}{2}} \cdot dv
<
\int_{v=0}^\infty \left( \frac{1}{\max(v,1)} \right)^{\frac{(1-\gamma)n-i+1}{2}} v^{\frac{(1-\gamma)n-i-r-1}{2}} \cdot dv
%=\\=
%\int_{v=0}^1 \hdots + \int_{v=1}^\infty \hdots
%=\\=
%\int_{v=0}^1 1 \cdot v^{\frac{(1-\gamma)n-i-r-1}{2}} dv +
%\int_{v=1}^\infty \left( \frac{1}{v} \right)^{\frac{(1-\gamma)n-i+1}{2}} v^{\frac{(1-\gamma)n-i-r-1}{2}} \cdot dv
%=\\=
%\int_{v=0}^1 \cdot v^{\frac{(1-\gamma)n-i-r-1}{2}} dv +
%\int_{v=1}^\infty v^{-\frac{r}{2} - 1} \cdot dv
=\\=
\frac{2}{(1-\gamma)n-i-r+1} + \frac{2}{r}
\leq
\frac{2}{(1-\gamma)n-t-r+1} + \frac{2}{r}
\end{multline}

%Note that had we replaced $\max(v,1)$ with $2 \max(v,1)$ we would have obtained a lower bound which is a factor $2^{-\frac{(1-\gamma)n-i+1}{2}}$ of the upper bound. Therefore the bound is tight at least up to this factor.

Combining (\ref{eq:D_i_derivation}), (\ref{eq:w_integral}) and (\ref{eq:v_integral_bound}) we obtain:
\begin{equation}\label{eq:D_i_bound}
D_i <
%c_1 \int_{w=0}^\infty  w^{\frac{n-i-1}{2}} e^{-w/2} dw \cdot \int_{v=0}^\infty \left( \frac{1}{v+1} \right)^{\frac{(1-\gamma)n-i+1}{2}} v^{\frac{(1-\gamma)n-i-r-1}{2}} \cdot dv
%<\\<
%\frac{1}{2^{(n-i+1)/2} \Gamma \left(\frac{r}{2} \right)  \Gamma \left(\frac{n-i-r+1}{2} \right)} 2^{\frac{n-i+1}{2}} \Gamma \left( \frac{n-i+1}{2} \right) \cdot \left( \frac{2}{(1-\gamma)n-t-r+1} + \frac{2}{r} \right)
%=\\=
\frac{\Gamma \left( \frac{n-i+1}{2} \right)}{\Gamma \left(\frac{r}{2} \right)  \Gamma \left(\frac{n-i-r+1}{2} \right)} \cdot \left( \frac{2}{(1-\gamma)n-t-r+1} + \frac{2}{r} \right)
\end{equation}

Since $L$ results in a rate loss of $\frac{1}{n} \log L$, and $\Gamma \left( \frac{n-i+1}{2} \right)$ is superexponential in $n$, we would like to express more explicitly the dependence on $n$. Using $\Gamma(t+1) = t \Gamma(t)$ with $t=\frac{n-t+1-2i}{2}$, $i=1,2,.. \lceil r/2 \rceil$ we can obtain the bound
\begin{equation}
\frac{\Gamma \left( \frac{n-t+1}{2} \right)}{\Gamma \left( \frac{n-t+1-r}{2} \right)} \leq \left( \frac{n}{2} \right)^{\lceil r/2 \rceil}
\end{equation}
therefore
\begin{equation}
D_i <
\frac{\left( \frac{n}{2} \right)^{\lceil r/2 \rceil}}{\Gamma \left(\frac{r}{2} \right)} \cdot \left( \frac{2}{(1-\gamma)n-t-r+1} + \frac{2}{r} \right)
\end{equation}
%\begin{multline}
%\Gamma \left( \frac{n-k+1}{2} \right)
%=
%\frac{n-k+1-2}{2} \Gamma \left( \frac{n-k+1-2}{2}  \right)
%= .. =\\=
%\prod_{i=1}^{\lceil r/2 \rceil} \frac{n-k+1-2i}{2} \Gamma \left( \frac{n-k+1-2{\lceil r/2 \rceil}}{2} \right)
%\leq
%\prod_{i=1}^{\lceil r/2 \rceil} \frac{n-k+1-2i}{2} \Gamma \left( \frac{n-k+1-r}{2} \right)
%<\\<
%\prod_{i=1}^{\lceil r/2 \rceil} \frac{n}{2} \Gamma \left( \frac{n-k+1-r}{2} \right)
%=
%\left( \frac{n}{2} \right)^{\lceil r/2 \rceil} \Gamma \left( \frac{n-k+1-r}{2} \right)
%\end{multline}
%Probably in a more detailed calculation the $\lceil \cdot  \rceil $ can be removed.
and
\begin{multline}
L = \prod_{i=1}^{t} D_i <
\frac{\left( \frac{n}{2} \right)^{t \lceil r/2 \rceil}}{\Gamma^{t} \left(\frac{r}{2} \right)}  \cdot \left( \frac{2}{(1-\gamma)n-t-r+1} + \frac{2}{r} \right)^{t}
=\\=
C_L \cdot n^{t \lceil r/2 \rceil}
\end{multline}

Substituting the above into (\ref{eq:Chernoff_error_bound}) proves Lemma \ref{lemma:pairwise_MIMO} for $\Lambda_x = \mt I$. The two assumptions on the parameters of the problem we have made in order for $L$ to be bounded are (a) $n \geq t + r$ which was needed in order that each new column of $\mt X$ would not be spanned by the previous columns and the columns of $\mt Y$ in probability 1, and (b) $\forall i \leq t: \gamma < 1 - \frac{i+r-1}{n} \Rightarrow \gamma < 1 - \frac{t+r-1}{n}$, is needed for the existence of $\{ D_i \}_{i=1}^t$.

Suppose now that $\mt X \sim \Normal^n(0,\Lambda_x)$. Using the Cholesky decomposition we can define a coloring matrix $\mt W$, $\mt W^T \mt W = \Lambda_x$ so that $\mt X = \mt W \cdot \mt X_w$ and $\mt X_w \sim \Normal^n(0,\mt I)$. Since by Property \ref{property2} the rate function is invariant to a linear transformation of $\mt X$ we would have $\Remp(\mt X_w, \mt Y) = \Remp(\mt X, \mt Y)$, therefore if Lemma \ref{lemma:pairwise_MIMO} holds with respect to the white signal $\mt X_w$ it also holds with respect to $\mt X$. With regard to the assumption that the columns of $\mt Y$ are linearly independent: if they are not, then the rate function is defined with respect to a smaller matrix $\mt Y'_{n \times r'}$ containing only the independent columns. Comparing with a full rank $\mt Y$, the random variables $d_i$ increase (i.e. $d_i' \geq d_i$) due to the smaller dimension of $\mt Y'$, therefore $L' \leq L$ and the lemma still holds.
\endofproof

\section{Comments and extensions}\label{sec:comments}
\textbf{Comparison with the SISO case:} Comparing Lemma \ref{lemma:pairwise_MIMO} with Lemma 4 of \cite{Individual_channels_full} for the SISO case $r=t=1$, which is proven by a direct calculation, the bound here is slightly worse due to the limitation $\gamma < 1 - \frac{t+r-1}{n} = (n-1)/n$ which stems from the use of the Chernoff bound.
%The reason is that in \cite{Individual_channels_full} the bound was directly calculated without using the Chernoff inequality. The lack of tightness is not related to the other inequalities used, since $D_i$ would diverge (not exist) for $\gamma = 1-1/n$. The loss due to this is rather small (in the order of 1 symbol).

\textbf{Comparison with MIMO capacity:}
The scheme above achieves the mutual information of a Gaussian MIMO channel but not its capacity. Achieving the capacity requires adaptation of the input distribution, which for the known AWGN channel $\vr y = \mt H \vr x + \vr v$ is performed by SVD and water pouring \cite{Telatar_MIMO}. The strength of the scheme is in the lack of any assumptions about the probability distribution, which make it applicable for example for non Gaussian noise or one that depends on the transmitted signal.

\textbf{Exploiting temporal correlation:}
In the current results, as in previous ones \cite{Individual_channels_full}, the rate function depends on the zero order empirical probability, and lacks the ability to exploit temporal correlation. However the results can be used to exploit such correlation in the SISO or MIMO channel, in a crude way, by applying the scheme on blocks of $k$ channel uses. The rate function over blocks is always superior to the single letter case, and the penalty is an increase in the fixed redundancy.

\textbf{Using empirical covariance instead of correlation:}
When the matrices $\hat {\mt R}$ in (\ref{eq:Remp_MIMO}) are replaced with the empirical correlation $\hat {\mt C}$ (where $\hat {\mt C}_{\mt X} \equiv n^{-1}(\mt X - n^{-1} \vr 1^T \mt X)^T(\mt X - n^{-1} \vr 1^T \mt X)$), the derivation is similar, except projection on an additional dimension (the all-ones vector) precedes the other projections. The results are the same with a loss of one dimension: $\gamma < 1 - \frac{t+r}{n}$ and $n > t+r$ are required, and there is a small variation in $C_L$.

\textbf{The complex MIMO channel:}
The results easily extend to the \textit{complex}-valued MIMO channel, using the same technique. The main difference is a double number of degrees of freedom in the derivation of $D_i$, which doubles the rate compared to Equation \ref{eq:Remp_MIMO}.

\textbf{Adaptivity:}
In \cite{Individual_channels_full}\cite{Individual_channels_isit} we presented a communication scheme using a low rate feedback, which dynamically adapts the transmission rate and achieves the rate functions
% presented there (the empirical mutual information for the discrete case and $-\half \log(1-\hat\rho^2)$ for the real valued case),
without outage. Such schemes are of higher practical interest. It is possible to show that the adaptive scheme of \cite{Individual_channels_full}\cite{Individual_channels_isit} achieves $\Remp$ of (\ref{eq:Remp_MIMO}) up asymptotically vanishing redundancy, and up to a set of $\vr x$ sequences having vanishing probability.
%Another important goal is finding an upper bound for the set of achievable rate functions. Additional research items appear in the previous papers \cite{Individual_channels_full}\cite{Individual_channels_isit}.

% --------------------------------------------- REFERENCES ------------------------------------------------

\end{document}